\documentclass[10pt]{iopart}
\usepackage{threeparttable}
\usepackage{graphicx}
\usepackage{xcolor}
\usepackage{listings}
\usepackage{siunitx}
\usepackage{url}

\begin{document}

\title[FPL\_VLSI]{Inductorless Fast Phase Logic: Enabling Two-Order-of-Magnitude Density Scaling for Superconductor VLSI
}

\author{Sasan Razmkhah$^1$, D. Scott Holmes$^{1,2}$, and Massoud Pedram$^1$}

\address{$^1$ SPORT Laboratory, Ming Hsieh Department of Electrical and Computer Engineering, University of Southern California, Los Angeles, California, USA}
\address{$^2$ IEEE IRDS}
\ead{razmkhah@usc.edu}
\vspace{10pt}
\begin{indented}
\item[]
\end{indented}

\begin{abstract} 
Fast phase logic (FPL) is a novel digital superconductor electronic (SCE) logic family that employs multiple junction types, including switching 0-Josephson junctions (0-JJs), non-switching 0-JJ stacks, and $\pi$-JJs. FPL enables flexible, automatable cell layouts, faster pulse propagation, reduced bias current via phase-shifting $\pi$-JJs, and minimized inductive loops, thereby reducing susceptibility to trapped flux and crosstalk. A fabrication process to support FPL is proposed. NbTiN superconductors offer small grain sizes, smooth surfaces, and thermal stability up to 400~$^\circ$C, while high-$J_c$, self-shunted JJs enable compact devices. AlN dielectrics provide good crystal matching to NbTiN, improving superconducting properties. Projections indicate that FPL, combined with the proposed process, can achieve a two-order-of-magnitude increase in integration density over conventional RSFQ logic and a five-fold reduction in supply current. The increased density reduces latency and improves computational throughput, while NbTiN-based devices provide higher output voltage and impedance, improving compatibility with CMOS circuits. Further fabrication advancements, such as higher-$J_c$ NbTiN-based JJs, higher processing temperatures, and stacked JJ structures, could enhance FPL implementation and scalability toward very large-scale integration (VLSI). FPL has the potential to significantly advance SCE technology, with near-term applications in accelerator cores for signal processing and artificial intelligence, and long-term potential in supercomputing. Its advantages are evaluated through an architectural study of a fast Fourier transform (FFT) circuit, with comparisons to CMOS and SFQ technologies.
\end{abstract}

\vspace{2pc}
\maketitle
\ioptwocol
\section{Introduction}
Superconductor electronics can achieve ultra-low-power, ultra-fast circuit operation by leveraging their macroscopic quantum properties. Below the critical temperature, Cooper-pair condensation leads to perfect diamagnetism and quantization of magnetic flux in superconducting loops. The resulting flux quantum,$\Phi_0 = \hbar / 2e \simeq 2.07\times10^{-15}$~Wb, arises from the $2e$ charge of Cooper pairs and provides the physical basis for both precision magnetometry and digital superconducting circuits \cite{razmkhah2023superconducting}. Flux quantization underpins Josephson-junction (JJ) based logic families, including single flux quantum (SFQ) \cite{likharev1991rsfq}, adiabatic quantum flux parametron (AQFP) \cite{takeuchi2013adiabatic}, and reciprocal quantum logic (RQL) \cite{RQL}, where information is encoded in discrete phase or flux states rather than voltage levels.

Josephson junctions supply the essential nonlinearity and switching mechanism in superconducting circuits. Governed by the DC and AC Josephson relations, a JJ undergoes a $2\pi$ phase slip when driven beyond its critical current, generating a quantized SFQ pulse \cite{likharev1991rsfq}. This phase-based switching enables picosecond-scale operation with switching energies on the order of $\Phi_0 I_c$, while superconducting interconnects support near-ballistic signal propagation. Despite these advantages, scalability remains limited by biasing complexity, wiring overhead, and integration density, motivating the exploration of new logic families.

Cryogenic temperature operation, although costly, is not the primary challenge for superconductor circuits. The fabrication infrastructure for superconductor circuits lags significantly behind CMOS, with few facilities currently equipped for modern superconductor fabrication \cite{tolpygo2020increasing}. Additionally, complex physics and the lack of accurate modeling make superconductor computing challenging \cite{razmkhah2024challenges,ceqip2023}. Circuit-level simulation tools, such as the Josephson Simulator (JSIM) \cite{JSIM} and the Portable Superconductor Circuit Analyzer (PSCAN) \cite{polonsky1991pscan}, use the resistively and capacitively shunted junction (RCSJ) model for SIS-type JJs under conditions where $T << T_c$. Projects like Supertools/ColdFlux have attempted to address tooling gaps with qPALACE for high-level design, JoSIM for circuit-level simulations, and TCAD/FLOOSS for process-level simulations \cite{fourie2023results}.

Despite these advances, current SCE technologies have substantial limitations. Implementing logic gates in superconductor electronics typically requires a higher number of circuit elements using two-terminal JJs compared to the three-terminal MOSFETs utilized in CMOS, which contributes to a significantly larger circuit area. Furthermore, traditional niobium (Nb) fabrication processes limit processing temperatures to below 200~$^\circ$C, which is substantially lower than the approximately 400~$^\circ$C limit typical of the CMOS back-end-of-the-line (BEOL). Consequently, the silicon oxide dielectric deposited on the chip has limited step coverage and gap-fill capability, severely limiting circuit density \cite{tolpygo2019advanced}.

Implementing logic gates typically requires more circuit elements using two-terminal JJs than three-terminal MOSFETs, which contributes to a larger physical size. SFQ logic gates are clocked, which limits logic depth per clock cycle and requires path-balancing elements. Most SFQ gates have a natural fanin/fanout of 1 \cite{razmkhah2024challenges}. Moreover, SCE lacks native memory with sufficient capacity or density. 

In CMOS, there are two complementary devices, PMOS and NMOS, for cell design. This enables compact logic implementation without large passive elements, providing a rail-to-rail voltage swing with minimal power consumption. However, SFQ logic implementations typically rely on a single type of JJ, requiring large inductors and resistors to emulate complementary behavior and resulting in physically large circuits and poor scalability. Introducing at least two JJ types with different intrinsic phase characteristics, such as 0- and $\pi$-JJs, enables complementary behavior in the phase domain. Logic using 0-JJs and $\pi$-JJs, designated as 0,$\pi$-JJ logic, enables implementation of logical functions without physically large devices, making VLSI feasible.

JJ's behavior depends on the device structure and materials, especially the barrier. The barrier material determines the type of JJ and its behavior in the circuit. The choice of barrier material (e.g., insulating, metallic, or magnetic) results in JJs with different voltage-phase and current-voltage behaviors \cite{golubov2004current}. Various types of JJs and their combinations can be explored to implement novel logic and circuits \cite{soloviev2021superconducting, cong2024superconductor}.

The most common fabrication technology for digital SCE is based on Nb superconductors and Nb/Al--AlO\textsubscript{x}/Nb junctions \cite{tolpygo2014fabrication}. Problems include Nb surface roughness and reactivity with oxygen or hydrogen, which degrades the superconducting properties, and instability of the thin ($\sim$1~nm) amorphous AlO\textsubscript{x} junction barrier. These problems limit fabrication processing and application temperature to below 200~$^\circ$C, which is much lower than the CMOS back-end-of-the-line (BEOL) temperature limits around 400~$^\circ$C. A consequence is that the SiO\textsubscript{x} on the chip is not of sufficient quality to fill small gaps with uniform material. The large grain size and roughness of Nb also cause increased parametric variability of JJ's electrical characteristics when the size decreases \cite{tolpygo2023progress}. These limitations result in either physically large circuits or large parameter variations. 

In this work, we build on the initial introduction of the fast phase logic (FPL) family in \cite{razmkhah2024high} by significantly advancing its practical applicability in very large-scale integration (VLSI). In addition to revisiting the core concepts of FPL, we introduce updated fabrication strategies for NbTiN and stacked 0-$\pi$ Josephson junctions, propose novel barrier materials to improve $\pi$-JJ performance, and develop a SPICE-compatible simulation methodology for accurate timing analysis.

Although the proposed FPL logic family is conceptually grounded in established Josephson physics and supported by previous demonstrations of $\pi$-JJ behavior and stacked-JJ fabrication, full device-level experimental validation remains an ongoing effort. Recent experimental work has verified the feasibility and performance benefits of $\pi$-JJs in SFQ circuits, as well as the fabrication of stacked high-$J_c$ nitride-based junctions and miniaturized self-shunted devices, providing evidence that the underlying device concepts required for FPL are physically realizable \cite{li2023superconductor,zhang2022fabrication,zheng2025nanoscale}. However, large-scale VLSI-capable fabrication processes for heterogeneous 0 and $\pi$ stacked junctions are still emerging \cite{lozano2024properties,herr2023scaling}. To address this, we are actively developing an NbTiN-based process to fabricate both 0 and $\pi$ JJs in stacked structures compatible with FPL cell design. The purpose of this work is therefore to motivate and quantify the architectural and circuit-level benefits enabled by such a process and to define the device- and circuit-level specifications required to guide fabrication development. To bridge the gap between device physics and circuit-level demonstration, we also introduce a modeling and simulation methodology that extracts device parameters and constructs SPICE-compatible models for realistic timing and performance analysis.

In the following sections, we present a complete system-level design framework, including a mixed clocking scheme, hierarchical placement strategies, and an automated layout tool. Finally, we demonstrate the architectural viability of FPL through a detailed case study of a 1024-point FFT circuit, highlighting its superior throughput, area efficiency, and potential for memory integration compared to CMOS and conventional SFQ technologies. Section 2 introduces the FPL family and outlines its advantages. In Section 3, we present various logic cells along with their characterizations. Section 4 discusses methods for large-scale implementation, whereas Section 5 presents the simulation results for several of these circuits. In Section 6, we study the FFT architecture to express the system-level benefits of FPL.

\section{Methodology}
\begin{figure}[htpb]
\begin{center}
\includegraphics[width=0.9\linewidth]{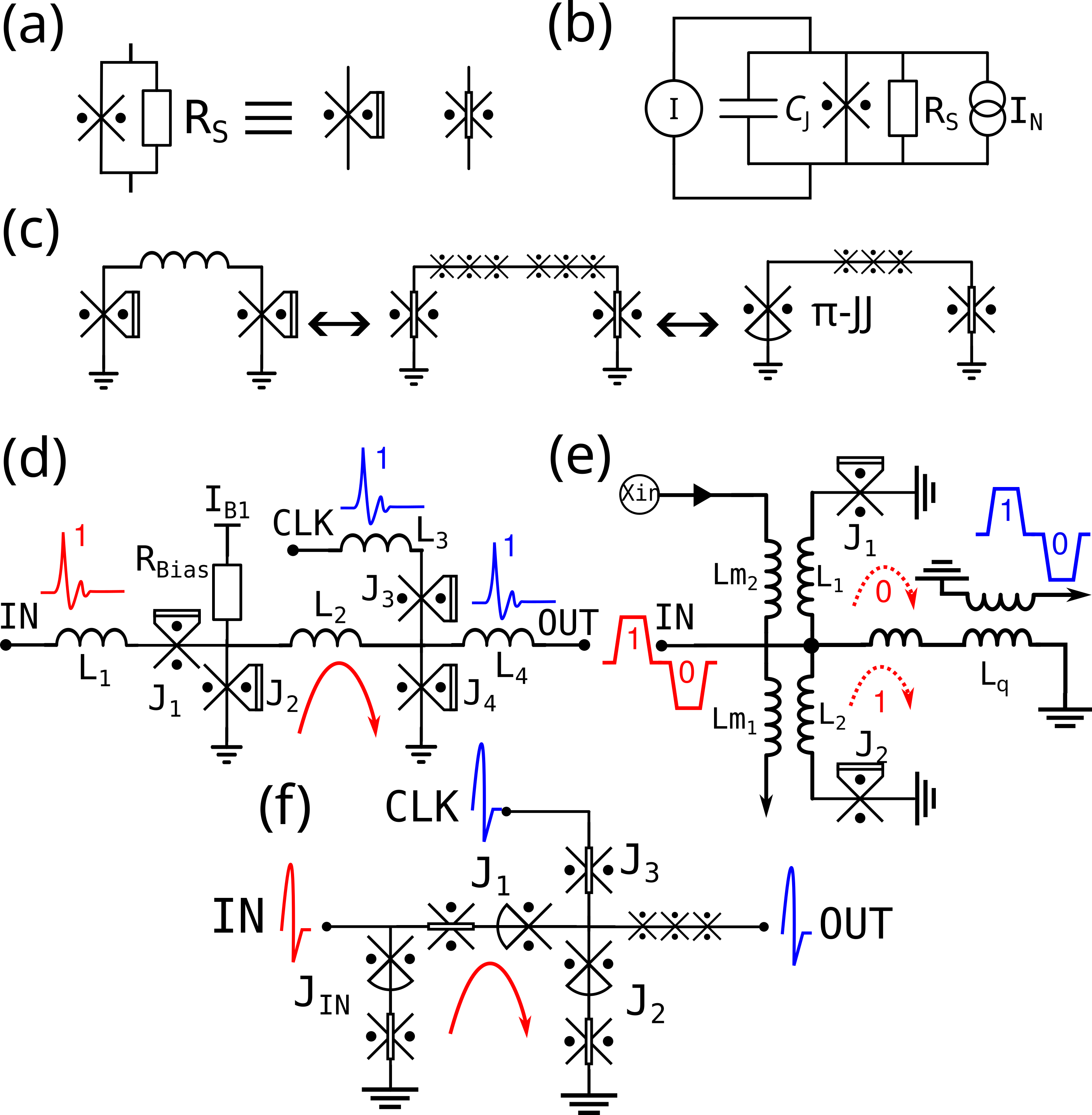}
\end{center}
\caption{Josephson junction and different Josephson technologies. (a) 0-JJ \cite{JJSymbolRef} with a shunt resistor and symbols for resistively shunted and self-shunted JJs, (b) RCSJ model of a 0-JJ with a noise source, (c) Storage loop formed from two JJs and an inductor or from combinations of JJs (d) Delay flip-flop (DFF) cell in SFQ technology; J2-L2-J4 form a storage loop that can store the incoming pulse and by the clock signal this pulse is released to the output, (e) Buffer cell in AQFP technology, in which input current polarity determines whether flux is in loop 0 (J1-L1-L$_q$) or in loop 1 (J2-L2-L$_q$) and the polarity of the output current, (f) FPL DFF cell; the storage loop here is formed by $J_IN$-J1-J2; when a clock pulse arrives and the loop contains a HFQ, J2 generates a pulse at the output.}\label{fig:1}
\vspace{-1em}
\end{figure}

RSFQ is one of the most widely studied and employed superconductor logic families. Although these circuits are fast, can operate at hundreds of GHz \cite{chen1999rapid}, can be power-efficient with an energy consumption of $I_c \Phi _0$ per switching action, which is equal to $\sim$~$10^3 k_B T \ln(2)$, and have a good parametric margin, they are not scalable to achieve very large-scale integration \cite{razmkhah2024challenges}. In RSFQ circuits, each switching JJ is resistively shunted, either externally or internally, to achieve nearly critical damping. The damping factor ($\beta _c$) is
\begin{equation}
    \beta _c = 2\pi \frac{R^2 C}{\Phi _0}
\end{equation}
\noindent where \textit{C} is the capacitance of JJ and $R = R_{shunt} || R_n \simeq R_{shunt}$. The circuit diagram and symbol are shown in Fig.~\ref{fig:1}(a), and the model of the shunted JJ is shown in Fig.~\ref{fig:1}(b). 

SFQ circuits require inductances, which are typically implemented as wire inductors in a superconductor layer. A storage loop can be used to store a cell's logical state. A superconducting loop with inductance $L$ and a JJ with critical current $I_c$ can store a flux quantum if $LI_c > \Phi_0$. Three methods for implementing a storage loop are shown in Fig.~\ref{fig:1}(c). RSFQ implements storage loops with a large inductor. The inductance of a wire inductor is small, especially when it is close to the ground planes, which contributes to large cell sizes. The inductance is also difficult to calculate because it depends on the wire's geometry and material properties as well as those of nearby components, making parametric cell design challenging. 

Moreover, the DC used to bias switching JJs results in a substantial current demand for the VLSI circuits. For example, an RSFQ 32-bit multiplier fabricated in the MIT LL SFQ5ee process requires 550~$mm^2$ of area and approximately 85~A of supply current, with almost all the current used to bias junctions to switch in a defined direction \cite{fourie2023results}. Approaches to reduce supply current include current recycling through a series of logic blocks \cite{semenov2019current} or AC distribution using transformers or capacitors \cite{vernik2014design}. Fig.~\ref{fig:1}(d) shows an RSFQ DFF cell. The diabatic quantum flux parametron (AQFP) circuits \cite{takeuchi2022adiabatic}, shown in Fig.~\ref{fig:1}(e), use an AC power supply and adiabatic switching to greatly reduce both the supply current and the dissipation of the power. However, scalability remains problematic due to the use of inductors and transformers \cite{ceqip2023}.

Another way to reduce the required supply current and switching energy is to decrease the critical current $I_c$ of the 0-JJs. A challenge with this approach is that the storage loop's inductance must be increased to maintain $LI_c > \Phi_0$. In low $I_c$ circuits with wire inductors, the inductor area is dominant and difficult to reduce.

\subsection{SFQ with Vertically Stacked JJ Inductors (VSJJIs)}
To address the circuit density limitations imposed by planar inductors in SFQ logic, the phase shift of JJs can be exploited as a compact replacement for geometric inductors. As shown in Fig.~\ref{fig:1}(c), the inductors in the storage loop can be eliminated by replacing the conventional SQUID loop with a series of 0-JJs in the loop, where the required phase bias is provided directly by the junctions. This approach enables a significant reduction in circuit area while preserving the required loop inductance for SFQ pulse storage and timing.

The total Josephson inductance of a junction with a sinusoidal current--phase relationship (CPR), $i = I_c \sin(\varphi)$, is given by~\cite{VanDuzer1999JJs} (Eq.~4.12(10)):
\begin{eqnarray}
L_J &=& L_{J0} \left[ \frac{\arcsin(i/I_c)}{i/I_c} \right]
= L_{J0} \left[ \frac{\varphi}{\sin(\varphi)} \right] \nonumber \\
    & & \approx \frac{L_{J0}}{\cos(\varphi)}.
\end{eqnarray}
where $L_{J0} = \Phi _0/(2\pi I_c)$, $I_c$ is the junction's critical current and $\varphi$ is the superconducting phase difference across the junction. The Josephson inductance varies from $L_{J0}$ at $i=0$ to $(\pi/2)L_{J0}$ at $i=\pm I_c$. In general, $L_J$ is nonlinear and frequency dependent and can diverge or become negative depending on the CPR and flux bias conditions. Note that the commonly used approximation $L_J \approx L_{J0}/\cos(\varphi)$ is valid only for small phase excursions or $(i/I_c)\stackrel{<}{\sim} 0.2$.

While the magnetic energy of a pulse propagates at a lower velocity in stacked JJs compared to the geometric inductor due to the higher capacitance of the JJs, the gates switch faster. For a gate to switch, the output JJ must pass the critical current, and in a stacked JJ, the current flows through a lower-impedance path due to the higher capacitance. This results in a faster switch at output JJs despite lower wave velocity in the stacked JJs themselves.

A significant advantage of Josephson inductors is that they can be vertically stacked to achieve higher inductance with less circuit area than traditional planar wire inductors~\cite{castellanos2019stacked}. In stacked Josephson junctions, multiple junctions are fabricated as multilayers and patterned into a single vertical pillar, connecting the junctions in series. The total inductance scales approximately linearly with the number of stacked junctions, while the occupied area remains nearly the same as that of a single junction. 

The value $L I_c$, important for SFQ circuit design, is the product of the stack inductance $L_{stk} \propto 1/I_{C~stk}$ and the critical current of the switching junction $L I_{C0}$, which is proportional to $I_{C0}/I_{C~stk}$. Scaling to smaller physical sizes or currents is easy, as both critical currents decrease together, and the ratio is a simple function of junction properties. Designing a circuit with wire inductors is much more difficult because inductance must increase as $I_c$ decreases and is highly dependent on geometry and material properties. 

Fabricating switching junctions from stacked inductor junctions has the advantage of making SFQ circuits more robust to process variations. A variation in $J_c$ cancels out, keeping the important $L I_c$ product constant. Junction diameter variations due to consistent underetching or overetching nearly cancel each other out. An alternative is to fabricate stacked inductor junctions separately from switching junctions, allowing the two $J_c$ values to be independent, albeit at the cost of additional fabrication steps.

\subsection{Fast Phase Logic (FPL)}

By combining stacked Josephson inductors with phase-engineered ($\pi$-shifted) loops, SFQ logic circuits can achieve substantially higher integration density while maintaining stable operation and robust timing margins. This is the basis for FPL.

The structure of an FPL circuit is shown in Fig.~\ref{fig:1}(f). JJ inductors have replaced the wire inductors. By replacing one of the JJs in a SQUID loop with a $\pi$-JJ, as shown in Fig.~\ref{fig:1}(c), a $\pi$-shifted loop is formed. The phase shift causes a constant current to develop in the loop. Proper component sizing allows the current generated by the pi-JJ to bias the switching 0-JJ in the loop. This is due to the quantization of the magnetic field in a superconductor loop that forces each loop to have a $2n\pi$ phase, where \textit{n} is an integer, and hence the loop will compensate for the $\pi$ shift in the phase by creating a current in the loop. 

Therefore, for a storage loop with an odd number of $\pi$-JJs, half a flux quantum is enough to store data in the loop, and its required loop inductance is reduced \cite{li2021low}. In a SQUID loop without and with $\pi$-JJ, the flux is calculated as
\begin{equation}
    \phi _{J1}+\phi _{J2}+\phi _{L} = 2n\pi, L = \frac{\beta _L \Phi _0 }{2\pi I_c} \simeq \frac{0.5\Phi _0}{I_c}
\end{equation}
\begin{equation}
    \phi _{J1}+(\phi _{J2} +\pi) +\phi _{L} = 2n\pi, L \simeq \frac{0.1\Phi _0}{I_c}
\end{equation}
\noindent where equations~3 \& 4 show the flux in normal and $\pi$-shifted SQUID loops, respectively. Here $\phi _{Ji}$ is the phase of the JJs, $\phi _{L}$ is the SQUID loops phase, and $\beta _L$ is the screening parameter, which in the SQUID loops used in logic is typically slightly greater than 1. Therefore, we can calculate the required inductance in each loop based on the bias applied to the JJs. 

\begin{figure}[htpb]
\begin{center}
\includegraphics[width=0.9\linewidth]{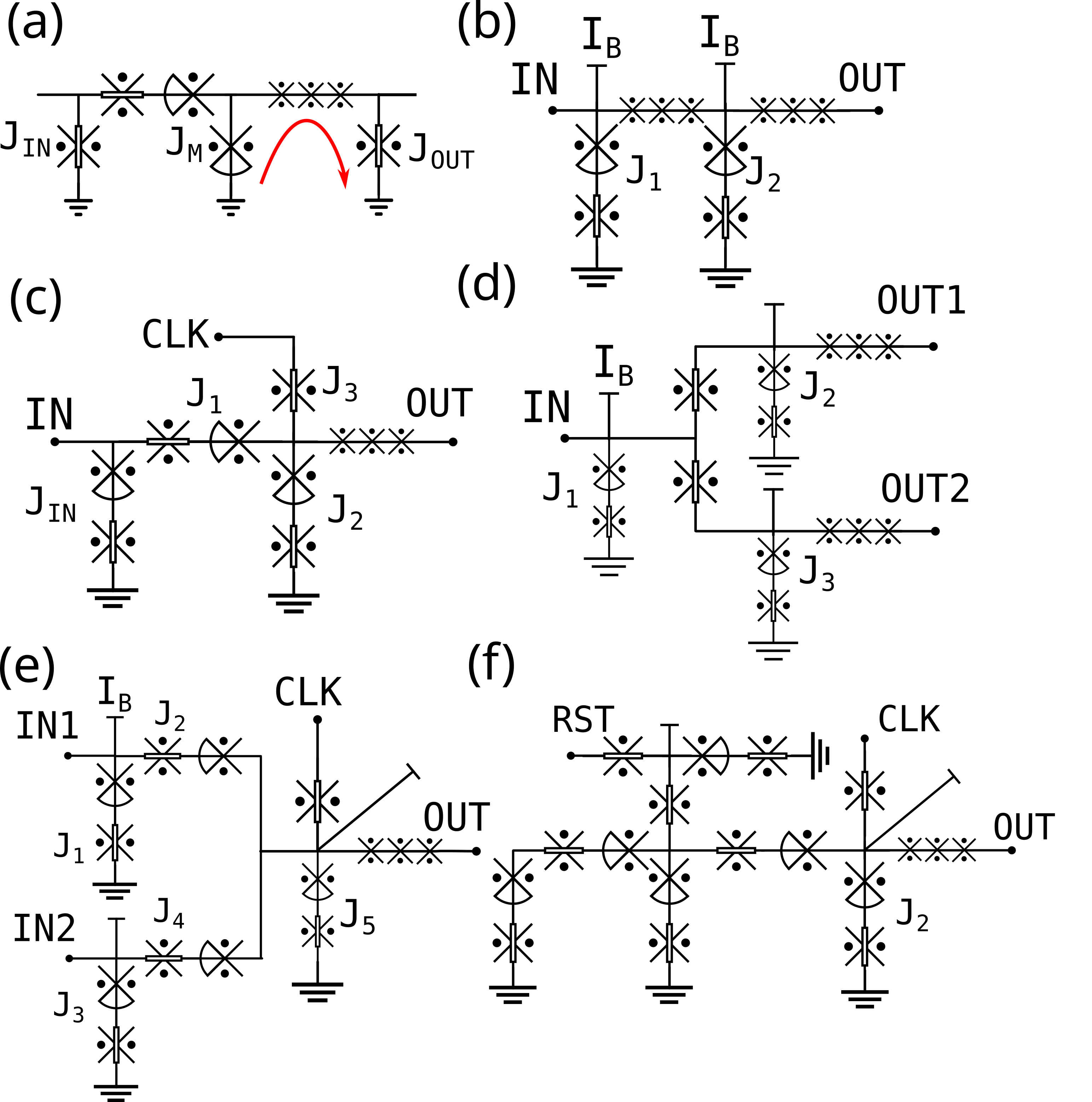}
\end{center}
\caption{FPL circuit and cell design: (a) Loops with odd numbers of $\pi$-JJs will have a circulating current, (b) Pulse propagation cell in FPL, (c) DFF cell that can also be used as shift register memory, (d) Splitter cell, (e) OR gate, (f) Non-destructive readout memory (NDRO) cell.}\label{fig:2}
\vspace{-1em}
\end{figure}

Moreover, using self-shunted JJs eliminates the need for shunt resistors on switching JJs. In \cite{yan2022intrinsically}, the authors demonstrated SFQ-based circuits using NbN/TaN/NbN JJs without shunt resistors. The key benefit of self-shunted JJ is its smaller area and ease of manufacture. In \cite{maksimovskaya2022phase}, the authors show that by combining SQUID loops with odd and even numbers of $\pi$-JJs (as proposed in Fig.~\ref{fig:2}(a)), we can see behavior similar to $2\phi$-JJs and thus implement phase logic. We combined this finding with high-$J_c$ JJs to implement a library of logic cells that can generate and propagate half-flux quantum pulses (HFQ) \cite{li2021low}. An HFQ is a quantized magnetic flux with the value equal to $\Phi_0 / 2$ and is generated by a bistable junction such as $2\phi$-JJ or their equivalent $\pi$ phase-shifted SQUID loops \cite{cong2024superconductor}.

In FPL cells, loops with an odd number of $\pi$-JJs are used as storage loops, which are used as a memory structure or to keep the cell state. On the other hand, loops with even numbers of $\pi$-JJs are used to propagate the generated pulse. Fig.~\ref{fig:2}(b) shows a Josephson transmission line (JTL) structure for pulse propagation. Fig.~\ref{fig:2}(c) is a DFF cell, which is the basic memory unit that we use for shift registers (SR), path balancing, and clocking. Fig.~\ref{fig:2}(d) is a datapath splitter that generates two pulses from one, to increase the fanout of logic cells. Fig.~\ref{fig:2}(e) shows the structure of an OR logic gate. When a pulse arrives from IN1 or IN2, it is stored in the loop involving J5; By the clock signal, this pulse is released to the OUT port. Fig.~\ref{fig:2}(f) proposes an NDRO cell design, a non-destructive memory unit used for random access memory implementation. The function and working principles of each of these circuits are discussed in more detail in \cite{razmkhah2024high}.  

\subsection{FPL Fabrication}
A robust fabrication process is needed to support the implementation of the proposed circuits using FPL technology. Problems with Nb superconductor-based fabrication processes described previously make it difficult to achieve significant improvements in circuit density. A superconductor with several advantages over Nb is delta-phase NbN. Still, it has a narrow composition range and is metastable, leading to non-uniformity across a substrate and between runs. NbTiN adds titanium to stabilize the structure, thereby improving the uniformity of the superconducting properties. NbTiN also has smaller grain sizes, smoother surfaces, and thermal stability up to 400~$^\circ$C, all improvements over Nb. NbTiN films 50 to 200~nm thick typically have $T_c$ in the 11 to 16~K range. The ability to fabricate NbTiN JJs with sufficiently low parameter variability remains to be proven.

To achieve optimal circuit performance, three junction types are employed: the 0-JJ, $\pi$-JJ, and stack 0-JJ. The 0-JJ serves as the switching junction and therefore requires precise control of its parameters for reliable circuit operation. The $\pi$-JJ provides the phase shift required for the switching elements, while the stack 0-JJ is used to realize the inductance in the superconductor loops. Because the stacked 0-JJ and $\pi$-JJ operate as non-switching elements, their fabrication tolerances are less critical than those of the switching 0-JJ. An additional advantage of using 0-JJs for switching and stacked 0-JJs as inductive elements is that a global variation in the 0-JJ barrier parameters, such as a shift in the $J_c$ value, affects both the critical current and loop inductance proportionally, keeping the product $L_{\text{loop}} I_c$ approximately constant and thereby improving circuit margins.

Furthermore, the maximum processing temperature can be increased to at least 400~$^\circ$C, allowing the deposition of higher-quality dielectrics using processes developed for CMOS fabrication. NbTiN has a higher kinetic inductance, enabling higher impedance for passive transmission lines. This will require a higher JJ impedance in the PTL driver circuits, which is already achieved by the smaller values $I_c$ used in this design. Barrier materials for self-shunted 0-JJs with NbTiN electrodes have included TaN \cite{zheng2025nanoscale, zhong2024single}, NbN \cite{Zhang2022NbNxJJs, Tolpygo2026NbNxJJs}, and Si \cite{Pokhrel2025AdvNbTiNCircuits}.

\begin{figure}[htpb]
\begin{center}
\includegraphics[width=1\linewidth]{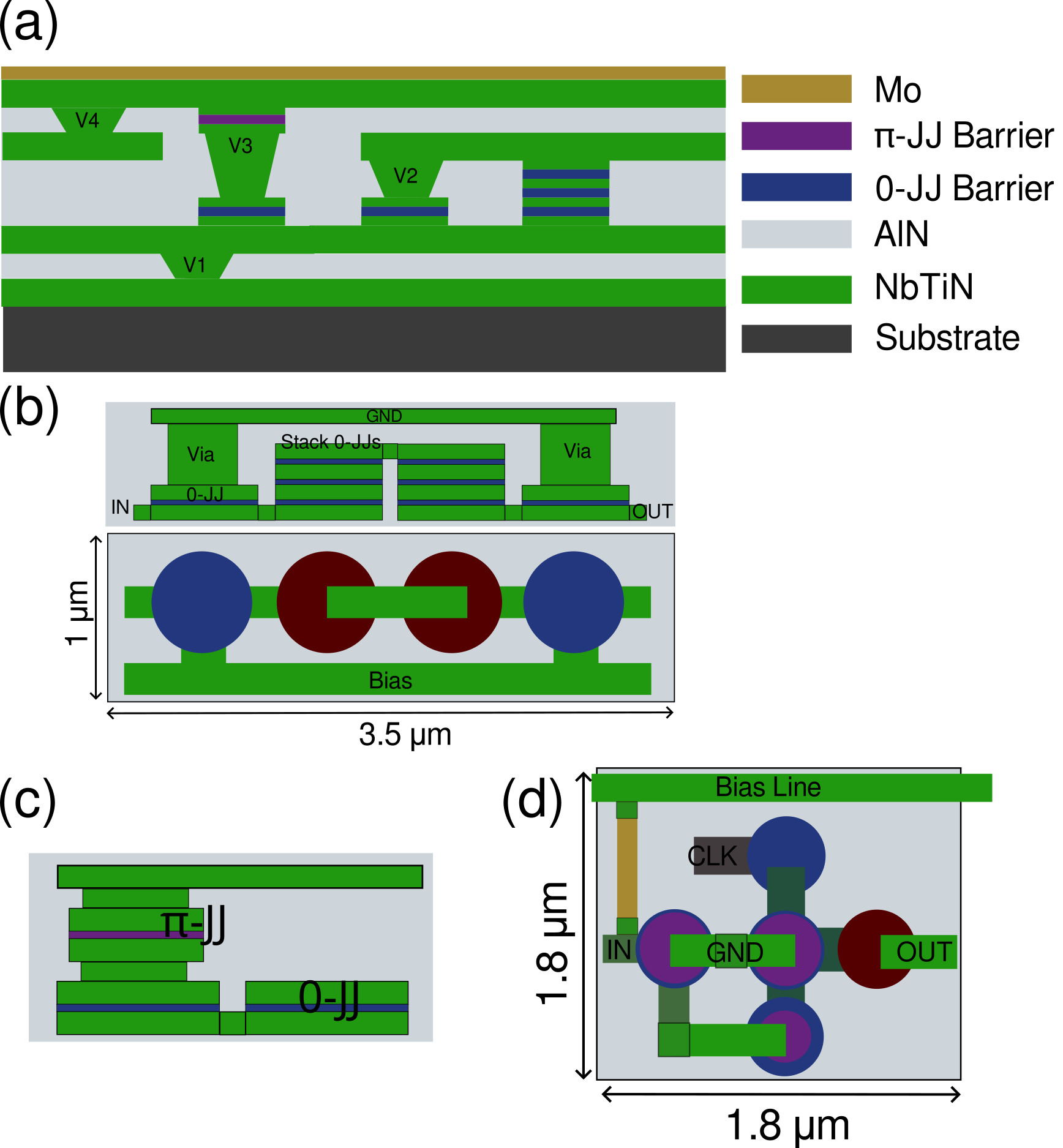}
\end{center}
\caption{(a) Proposed fabrication process stackup cross section, (b) JTL cell, (c) 0,$\pi$ switching stack connected to a 0-JJ, (d) DFF cell top view.}\label{fig:3}
\vspace{-1em}
\end{figure}

Fig.~\ref{fig:3}(a) shows the proposed fabrication stackup to achieve high-density superconductor integrated circuits. Junction barriers with appropriate resistivity allow fabrication of high $J_c$, self-shunted junctions with non-hysteretic switching behavior that do not require external shunt resistors. Additionally, these 0-JJs can be stacked to provide greater inductance in a compact device. The thickness of the superconductor between barrier layers should be greater than the coherence length $\xi$, which is only about 5~nm for NbTiN. 
For the $\pi$-JJ barrier material, ferromagnetic materials such as Ni, NiCu, and PdNi have been used with some success \cite{birge2024ferromagnetic, Yamashita2020PiPhaseShifter, Takeuchi2024NbNbasedHalf, Pham2024NbNbasedTunnel}. However, $\pi$-JJs have proven difficult to produce reliably or economically for VLSI applications using such ferromagnetic barrier materials. Problems include overly strong magnetization that requires very thin barrier layers, sensitivity to interfacial roughness and compositional variations, multiple domains that reduce $\pi$-JJ yield (e.g., Ni), high cost (e.g., PdNi), or crystal-structure mismatch with the superconductor electrode material. Alternative barrier materials considered include ferrimagnets  \cite{madden2022ferromagnetic} and altermagnets \cite{lu2024phiJJ}. 

Fig.~\ref{fig:3}(b) shows a JTL cell center loop with two switching JJs connected by a ground plane and by two stacks of non-switching 0-JJ compact inductors, (c) shows a 0,$\pi$ switching stack connected to a 0-JJ, and (d) shows a DFF cell layout top view.

\subsection{Simulation Methodology}
\label{Sec:SimMethod}
The proposed devices were simulated using the basic RCSJ model in JoSIM \cite{delport2019josim}. 
\begin{lstlisting}[language=python, basicstyle=\ttfamily\footnotesize]
.MODEL JJ(RTYPE, VG, DELV, R0, RN, CAP, 
          ICRIT, PHI,CPR)
\end{lstlisting}
\noindent where \textit{RTYPE} specifies whether to use the piecewise linear resistor model for normal currents, \textit{VG} is the gap voltage, \textit{DELV} is the transition voltage from subgap to normal, \textit{R0} is the subgap resistance, \textit{RN} is the normal resistance, \textit{CAP} is the capacitance per $\mathrm{\mu m}^2$, \textit{ICRIT} is the critical current density, \textit{PHI} is the CPR phase shift (for $\pi$-JJ models), and \textit{CPR} is the current-phase relation harmonic amplitudes (for $2\phi$-JJ models). For our simulations, we used this model to model 0- and $\pi$-JJs, using estimated values for \textit{VG}, \textit{R0}, \textit{RN}, \textit{CAP}, \textit{ICRIT}, and \textit{PHI} based on the proposed fabrication process and device designs. While this model is not sufficiently accurate for $\pi$-JJs \cite{zhong2024single}, it will provide a baseline for circuit operation. We improved this model in Listing~\ref{lst:jj_RCSJ}. 

\subsection{Parameter Calculation}
The Josephson junction (JJ) parameters used in our SPICE/JoSIM simulations are derived from standard superconducting relations, based on target material specifications for NbTiN-based junction stacks. The quantities $I_c$, $R_N$, $R_0$, $C$, $V_G$, and $f_c$ are computed from the following closed-form expressions, where the energy gap is calculated as $\Delta \approx 1.76 \, k_B T_c = 1.8$~meV. The critical current density $J_c$ is determined by quantum tunneling of Cooper pairs through the insulating barrier and is therefore highly sensitive to the barrier thickness $d_b$ and effective barrier height $\phi_b$. Using a simplified WKB tunneling approximation, the dependence of $J_c$ on barrier properties can be expressed as

\begin{equation}
    J_c \propto \exp\!\left(-2 d_b \sqrt{\frac{2m\phi_b}{\hbar^2}}\right),
\end{equation}

\noindent where $m$ is the electron mass and $\hbar$ is the reduced Planck constant. This exponential dependence implies that nanometer-scale variations in $d_b$ or small changes in $\phi_b$ can alter $J_c$ by orders of magnitude, highlighting the importance of precise barrier engineering. For NbTiN-based SIS junctions, a commonly used practical approximation is:

\begin{equation}
    J_c \simeq 3.16\times10^{10}\,\frac{\sqrt{\phi_b}}{d_b}\,
\exp\!\left(-1.025\,\sqrt{\phi_b}\,d_b\right),
\end{equation}

\noindent with $\phi_b$ in eV and $d_b$ in nm. Assuming the barrier height in TaN with (111) surface that is compatible with NbTiN is 2.30 eV, and the barrier thickness is 2.28~nm, we get about $0.6\,\mathrm{mA}/\mathrm{\mu m}^2$ for the critical current density of the junction. Using representative material and device assumptions of $T_c = 12\,\mathrm{K}$, $J_c = 0.6\,\mathrm{mA}/\mu\mathrm{\mu m}^2$, $A = 1\, \mathrm{\mu m}^2$, $C_A = 70\,\mathrm{fF}/\mathrm{\mu m}^2$, $r_{sg}=10$, and $\Phi_0 = 2.07\times 10^{-15}\,\mathrm{Wb}$, we obtain the parameters for simulation.

For an SIS 0-JJ using NbTiN/TaN/NbTiN materials, the values for \textit{VG}, \textit{R0}, \textit{RN}, \textit{CAP}, \textit{ICRIT}, and \textit{PHI} are 3.6 mV, 140 $\Omega$, 14 $\Omega$, 70~fF/$\mathrm{\mu m}^2$, 0.6~mA/$\mathrm{\mu m}^2$, and 0, respectively. For the switching element, we assumed an SFsIS 0 and $\pi$-JJ stack with NbTiN/Ferri/NbTiN/TaN/NbTiN materials. These values are 3.6 mV, 20 $\Omega$, 8 $\Omega$, 40~fF/$\mathrm{\mu m}^2$, 1 mA/$\mathrm{\mu m}^2$, and $\pi$, respectively.

For $\pi$-junctions, the same parameter computation applies with the intrinsic phase shift enforced in SPICE using $I = I_c \sin(\varphi + \pi)$. Parameter sweeps of $\pm 20\%$ confirmed that the relative performance trends and scaling conclusions of the proposed FPL family remain unchanged.

Another option is to implement the models in a SPICE simulator. We chose NGSPICE because it is open source and used in the KiCAD software package. In SPICE, a JJ can be modeled as a phase-dependent current source described by the DC Josephson effect $I(t) = I_c sin[\varphi (t) + \theta]$ where the $\theta$ is the phase shift intrinsic to the JJ that can determine if it is a 0- or $\pi$-JJ. The phase of the JJ $\varphi (t)$ is calculated using the AC Josephson effect from the voltage of the JJ as $\varphi (t) = 2e/\hbar \int V(t)$. With these equations and functions for the subgap and normal resistance, we can model the JJ with the sub-circuit in Listing~\ref{lst:jj_RCSJ}.

In the JJ model, `Cphi' integrates `Gphi' to compute the phase of the JJ based on the AC Josephson effect, `Bjj' is the current source dependent on the phase difference, `Cjj' is the JJ capacitance, and `Bpq' models the quasi-particle current using a piecewise linear resistive model with subgap resistance `Rsub' and normal resistance `RNorm'. The `area' parameter determines the JJ size, and `phase\_shift' determines its type.

\begin{lstlisting}[language=python, caption={NGSPICE sub-circuit model for 0- and $\pi$-JJs},label={lst:jj_RCSJ}, basicstyle=\ttfamily\footnotesize]
.subckt jj_RCSJ n1 n2 area=1 phase_shift=0
.param icrit = {area * ic0}
.param cj = {area * cj0}
Gphi 0 phi VALUE = {2*'PI'/phi0 * V(n1,n2)}
Cphi phi 0 1
Rleak phi 0 1k
Bjj n1 n2 I = {icrit * sin(V(phi) 
    + phase_shift)}
Cjj n1 n2 {cj}
Bqp n1 n2 I = {V(n1,n2) / (Rsub+0.5*(RNorm-Rsub) 
    * ((abs(V(n1,n2))-(Vg-delta)) 
    / delta+sgn(abs(V(n1,n2)) - (Vg-delta))) 
    * (1-sgn(abs(V(n1,n2))-(Vg+delta)))/2 
    + (RNorm - Rsub) * (1+sgn(abs(V(n1,n2)) 
    - (Vg+delta)))/2)}
.ends jj_RCSJ
\end{lstlisting}

\begin{figure}[htpb]
\begin{center}
\includegraphics[width=1\linewidth]{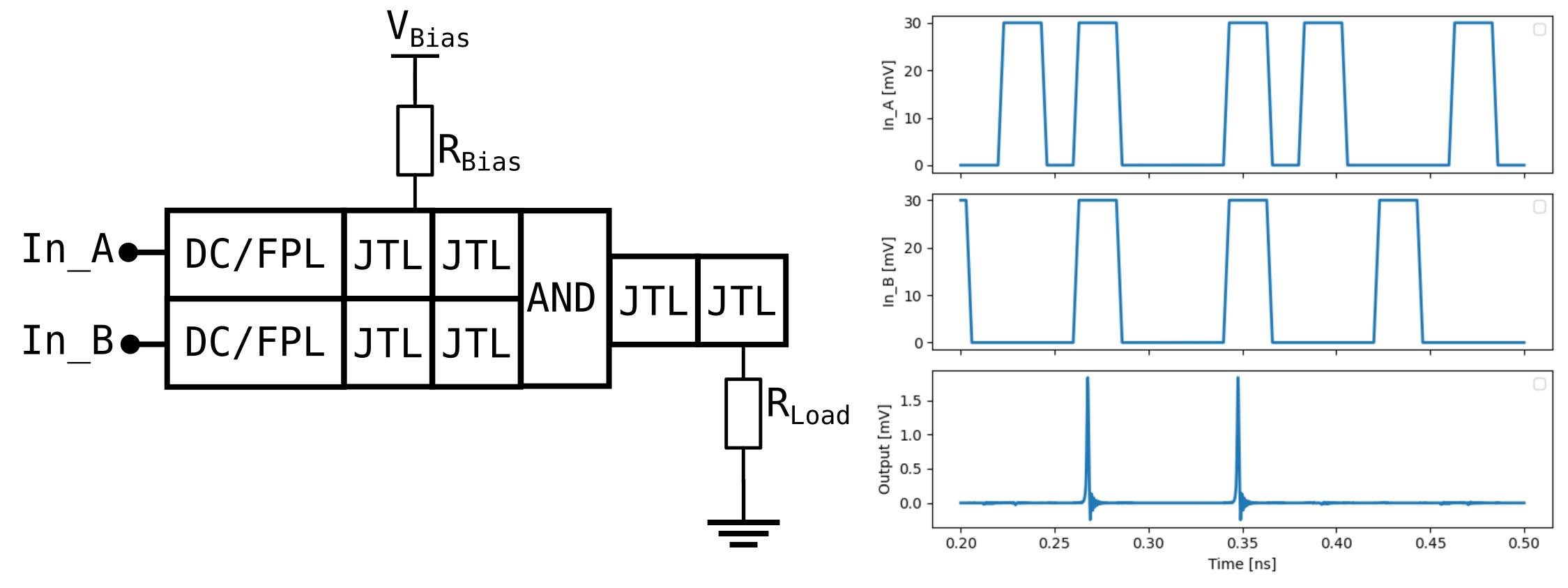}
\end{center}
\caption{FPL asynchronous AND gate schematic and SPICE simulation results. The inputs are applied to a DC/FPL cell over a 50~$\Omega$ resistor. The DC/FPL converters trigger on the rising edge of the input and generate flux-quantized pulses whenever the voltage crosses a defined threshold. The gate operates on a threshold principle, producing an output pulse when the combined input currents exceed the critical current of the output JJ, with a permissible input arrival skew of only a few picoseconds.}\label{fig:4}
\vspace{-1em}
\end{figure}

Simulation results for a high-$J_c$ JJ stack in an AND gate testbench are provided in Fig.~\ref{fig:4}. The test bench features two DC inputs that are converted to pulses via DC/FPL converters and then fed to an AND gate via JTLs. The output is read from a load resistor. A 0.5 ps delay is observed between the inputs and the output of the AND gate, which is ten times faster than that for conventional RSFQ circuits.

\subsection{Large Scale Implementation}
A suitable logic family and a robust fabrication process are essential for achieving superconductor-based VLSI circuits and systems, along with an architecture compatible with the chosen logic family and a software tool suite to optimize and map various target architectures and circuits to it. Pulse-based logic operates on the principle that the presence of a pulse represents a logical `1,' while its absence represents a logical `0.' Consequently, a clock signal is required to distinguish a logical `0' from the absence of data, resulting in a highly pipelined structure in SFQ architectures. SFQ architectures also require an extensive clock distribution network to deliver the clock signal accurately to each cell. Cells must be path-balanced to ensure data signals align with the clock signal. Current SFQ implementations employ either a clock-follow-data approach with path-balanced cell-delay-based path balancing or a zero-skew clock distribution approach. The former is unreliable for large circuits due to jitter-induced data loss, while the latter introduces significant overhead due to the need for path-balancing DFFs \cite{fourie2023results}.

The integration capability and speed of FPL circuits make them well-suited for realizing superconductor-based VLSI. For implementing an FPL architecture, we propose a mixed clocking scheme that utilizes macrocells with clock-follow-data functionality and an H-tree clock network between macrocells for zero-skew clock distribution \cite{dejima2020layout}. The hybrid JTL-PTL flow integrates hierarchical design principles, focusing on optimizing global path balance and reducing full path balancing (FPB) overhead while preserving overall design integrity. By selectively removing FPB at specific logic levels, configurable as a hyperparameter, the flow maintains the global path balance. Clusters are initially formed using fan-in-cone partitioning, followed by the creation of macrocells through clustering functions. The size of these macros, governed by another hyperparameter, influences their complexity and the overall number of macros. Intra-level placement prioritizes the positioning of signal cells and the construction of clock trees within each macrocell. By using logic-level-based, detailed placement, we enhance routing efficiency, which is particularly suited to concurrent-flow clock methodologies. We also apply macro-info propagation for subsequent inter-level placements.

Inter-macro placement further optimizes macro positions and clock-tree structure to minimize half-perimeter wire length (HPWL) and overall placement area, while ensuring H-tree balance. The flow incorporates adjustable clock tree synthesis (CTS) skew within macros post-routing, achieved through dedicated JTL cells, enabling precise timing adjustments both intra- and inter-macro. This comprehensive approach supports stage-specific optimization objectives, enabling iterative design refinements throughout the design process and thereby enhancing overall performance and manufacturability.

\begin{figure}[htpb]
\begin{center}
\includegraphics[width=0.9\linewidth]{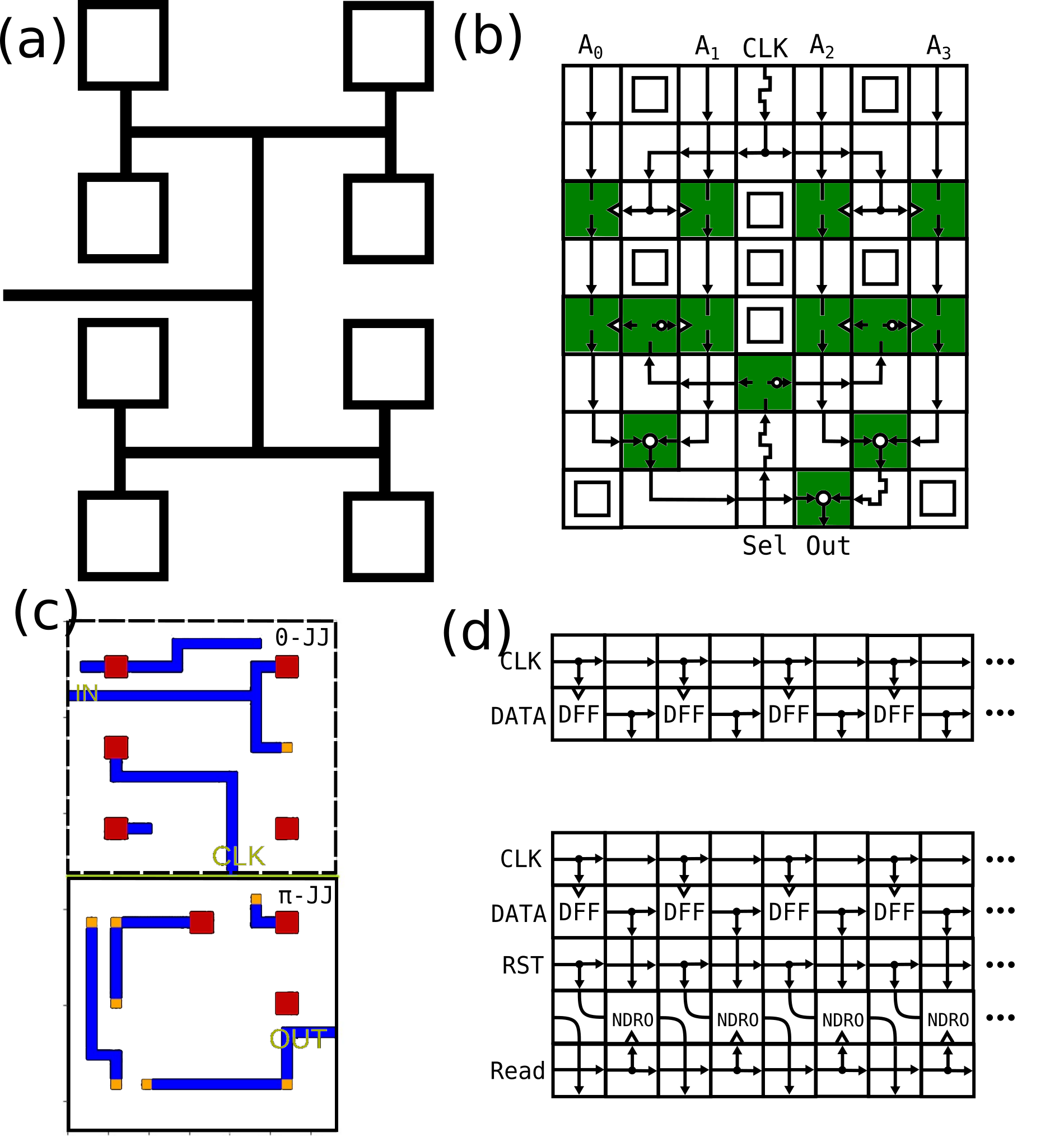}
\end{center}
\caption{FPL VLSI: (a) A zero skew clocking with H-tree for clock distribution between macrocells, (b) Macrocell (4-1 parallel to serial converter) implementation with the JTL delay lines, the solid green cells are logic, and the rest are splitter and active transmission lines, (c) PhaseConnect tool that we developed for automated cell layout design and optimization, (d) shift register and random access memory design for pulse-based technology}\label{fig:5}
\vspace{-1em}
\end{figure}

Fig.~\ref{fig:5}(a) illustrates an H-tree clock network that links the macrocells. Fig.~\ref{fig:5}(b) shows the output of our tool, which routes small circuit partitions using JTL lines, with path balancing and clocking managed through JTL delays. Since FPL is designed without geometric inductors and circuit behavior is primarily determined by Josephson junctions, its circuit layout and fabrication can follow a methodology similar to CMOS processes, enabling a more standardized, scalable design flow. In contrast, the RSFQ logic relies on inductors that are inherently formed by the physical geometry of superconducting interconnects. These inductors must be carefully designed and optimized using programs such as Inductex to ensure precise circuit operation, making the RSFQ layout highly dependent on electromagnetic simulations. This fundamental difference makes FPL more adaptable to automated circuit design and integration while simplifying fabrication constraints compared to RSFQ. Fig.~\ref{fig:5}(c) shows our tool, PhaseConnect \cite{Li2024PhaseConnect}, which automatically optimizes and generates the layout of the FPL circuit from a netlist by arranging 0- and $\pi$-JJs on a grid and routing connections between them.

One of the primary challenges in SCE-VLSI is achieving high-density memory integration. In \cite{tanaka2015development}, Tanaka et al. designed a processor core with only 256 bits of random access memory (RAM) and a chip area of 1.32~mm $\times$ 1.95~mm. Fig.~\ref{fig:5}(d) shows two types of memory frequently used in logic circuits. The shift register memory is faster and more compact, but is less flexible, making it less suitable for processor architectures like RISC-V. For the FPL shift register memory, each bit requires an area of 6~\si{\micro\meter\squared} with overhead. For RAM, NDRO cells, which are larger than DFFs, are needed. In FPL technology, each RAM bit occupies about 16~\si{\micro\meter\squared} of the area, including overhead. This results in a total area of 4096~\si{\micro\meter\squared} for 256-bit RAM.

\subsection{FPL-CMOS Co-design}
Although FPL cells are significantly denser than conventional SFQ circuits, the memory implementation in FPL remains nearly two orders of magnitude larger in size compared to its CMOS counterparts, as shown in Table~\ref{tab:Cell_Area_Comparison}. Similarly, the area estimates reported in this table for FPL cells are layout-driven physical projections generated using the same placement methodology implemented in our PhaseConnect tool suite, based on stacked-junction footprints, rather than experimentally fabricated chips. The junction dimensions, layer counts, and minimum feature spacing are derived from our target NbTiN fabrication flow and associated mask set, rather than existing foundry-qualified PDK values. The RSFQ cells are based on our ColdFlux cell library, which was validated for the MIT LL SFQ5ee fabrication process.

\begin{table}[htbp]
\centering
\begin{threeparttable}
\caption{Comparison of representative cell areas for combinations of logic implementation (RSFQ, SFQ, FPL, CMOS) and fabrication technology (SFQ5ee, 0,stack, 0,$\pi$,stack, FinFET 12~nm). MIT LL SFQ5ee technology has only 0-JJs. 0,stack technology has both 0-JJs and vertically stacked Josephson junction inductors (VSJJIs). 0,$\pi$,stack technology also has $\pi$-JJs. RSFQ values are based on published process data. FPL values are projection-based estimates generated using assumed stacked-JJ dimensions that have not yet been demonstrated.}
\label{tab:Cell_Area_Comparison}
\small
\begin{tabular}{|p{1cm}|p{1.3cm}|p{1.3cm}|p{1.3cm}|p{1.2cm}|}
\hline
\multicolumn{1}{|c|}{} & \multicolumn{4}{c|}{\textbf{Area ($\mu$m\textsuperscript{2})}} \\ \cline{2-5}
\textbf{Cell} & \textbf{RSFQ SFQ5ee} & \textbf{SFQ \mbox{0,stack}} & \textbf{FPL \mbox{0,$\pi$,stack}} & \textbf{CMOS 12~nm} \\ \hline \hline
JTL & 625 & 3.5 & 0.65 & -- \\ \hline
DFF & 625 & 12 & 2.4 & 0.4--0.8 \\ \hline
Merger & 625 & 10 & 2 & -- \\ \hline
Splitter & 625 & 8 & 1.8 & -- \\ \hline
OR & 2500 & 15 & 2.95 & 0.25--0.4 \\ \hline
AND & 2500 & 15 & 2.7 & 0.2--0.35 \\ \hline
NDRO & 2500 & 15 & 3 & *0.03 \\ \hline
\end{tabular}
\begin{tablenotes}
\item[*] 6T-SRAM memory cell
\end{tablenotes}
\end{threeparttable}
\end{table}

This presents challenges for architectural scalability and flexibility. To overcome this limitation, we explored alternative memory technologies operable at cryogenic temperatures, including bistable vortex memory (BVM)~\cite{karamuftuoglu2024superconductor}, cryo-CMOS~\cite{saligram2024future}, and memristors~\cite{song2023recent}.

Among these, BVM offers the highest speed and lowest power consumption; however, its density is still much lower than cryo-CMOS or memristor-based memory arrays. Memristors become more stable at cryogenic temperatures, offering non-volatility, fast switching, excellent scalability, and the highest density among these technologies. They operate reliably over a broad temperature range. Although the read operation of memristors is highly energy efficient, the write operation requires elevated voltages and energy at low temperatures, and device-to-device variability increases~\cite{alam2023cryogenic}. Cryo-CMOS memory, although not as dense as memristors, benefits from advanced fabrication processes that yield highly predictable device behavior. Moreover, at cryogenic temperatures, both the leakage current and the operating voltage are significantly reduced~\cite{saligram2024future}.

Therefore, for read-only memory applications such as storing twiddle factors in FFT architectures, memristors are an ideal choice due to their one-time programming and low-power read operations. For high-density memory needs, cryo-CMOS presents a practical solution, while inductorless BVM is well suited for low-density, high-speed memory blocks. 

\begin{figure}[ht!]
\begin{center}
\includegraphics[width=0.9\linewidth]{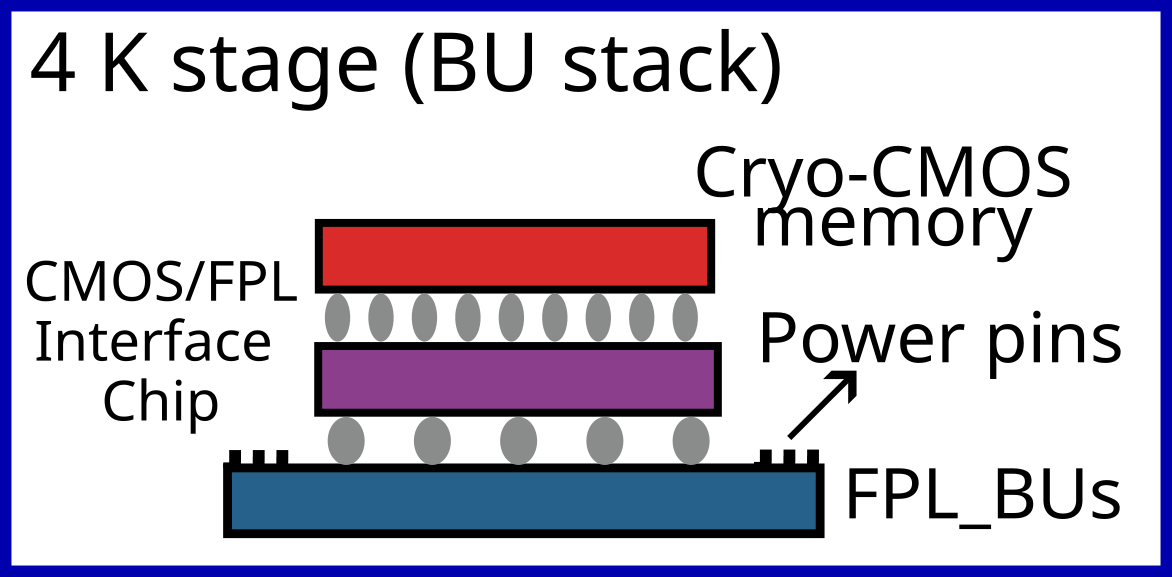}
\end{center}
\caption{Our proposed stacking method is based on the MCM and flip-chip bonding the CMOS memory as the TW weights to the interface chip (containing the parallel to serial converters), and the FPL BU chip with BVM feedback memory on it. }
\label{fig:7}
\vspace{-1em}
\end{figure}

In this work, due to the limited memory size, only cryo-CMOS memory with some on-chip superconductor memory for buffering would be sufficient. Integration is illustrated in Fig.~\ref{fig:7}, where cryo-CMOS serves as the storage for the TW constants, BVM for the feedback memory, and an interface chip translates CMOS clock and logic levels to FPL’s pulse-based format while preserving high performance through serialized multi-line inputs.

\subsection{Architecture Study}
\begin{figure}[htpb]
\begin{center}
\includegraphics[width=1\linewidth]{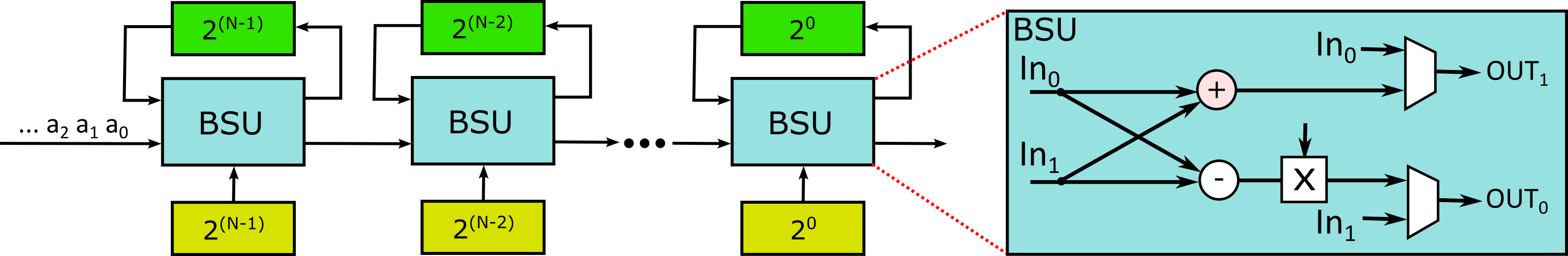}
\end{center}
\caption{Architectural block diagram of an SDF FFT circuit and each butterfly skip unit (BSU). BSU (cyan block) performs and skips butterfly operations, the feedback memory units (green blocks) store intermediate results for feedback, and the coefficient memory units (yellow blocks) store twiddle factors. For SFQ implementations, these memories can be implemented with shift registers. }
\label{fig:6}
\vspace{-1em}
\end{figure}

To investigate the potential advantages of FPL circuits, we designed a comparison circuit using combinations of logic implementation (RSFQ, SFQ+VSJJIs, FPL, CMOS) and fabrication technology (SFQ5ee, proposed 1, proposed 2, FinFET 12~nm). The single-path delay feedback (SDF) fast Fourier transform (FFT) architecture, a highly efficient FFT architecture, was selected for comparison \cite{ingemarsson2017efficient}. The architecture block diagram is shown in Fig.~\ref{fig:6}. This FFT architecture is particularly effective in streaming data systems. It is made up of a series of butterfly units interleaved with simple memory elements, such as RAM, which can be replaced with delay lines or shift registers. A key optimization in our design is the integration of butterfly skip units, which dynamically bypass certain computation stages based on FFT stage requirements, reducing latency and improving throughput in specific configurations. The streaming nature of SDF enables deeply pipelined operations that align exceptionally well with the characteristics of superconductor digital logic, such as SFQ circuits, which natively support high clock frequencies and fine-grain pipelining.

The selection of the fast Fourier transform (FFT) circuit, specifically the single-path delay feedback (SDF) architecture, was a deliberate choice, not to bias the results, but to highlight the alignment between the core strengths of FPL and a representative and performance-critical application in superconductor electronics. The SDF FFT is inherently compatible with deeply pipelined, stream-based computations, where FPL and other superconductor logic families, such as SFQ, excel. In deeply pipelined superconductor electronics, stream-based computation is significantly more efficient than instruction-based sequential execution, as found in architectures like RISC-V, which require instruction fetch, decode, and execution stages that introduce control complexity, large RAM arrays, and additional latency. In contrast, the regular, deterministic data flow of the SDF FFT enables continuous processing with minimal control overhead, simplifying control logic, ensuring predictable latency, and maximizing bandwidth utilization—characteristics crucial for real-time cryogenic systems and applicable to most combinational circuits. 

These architectural advantages align closely with the computational demands of modern artificial intelligence and real-time signal processing. Deeply pipelined superconductor electronics can process deterministic data streams far more efficiently than the instruction-based sequential execution found in traditional processors. Moreover, the minimal memory requirements of the deeply pipelined SDF architecture can be fully met by shift registers rather than physically larger random-access memory. Therefore, the FFT architecture was chosen because it shows FPL’s advantages in a domain-aligned setting. The methods and innovations introduced in this work provide general-purpose foundations for scaling FPL to a wide range of computational workloads, including integer arithmetic, AI acceleration, and scientific computing.

Scalability is another advantage of the FFT architecture. As the FFT size increases, the architecture scales linearly by adding more stages, each consisting of a butterfly unit and a corresponding memory element. This modular nature makes the design highly adaptable to different problem sizes without requiring a complete redesign. Furthermore, the memory requirements in SDF are minimal; each stage requires only a small buffer, which can be implemented using simple shift registers (SR) or circular shift registers (CSR). These memory structures are much more practical to implement in superconductor logic compared to RAM, which is both area- and power-intensive at cryogenic temperatures. Shift registers, in contrast, can be implemented efficiently using chains of SFQ D-flip-flops, making the SDF FFT a natural fit for high-speed, energy-efficient superconducting computation platforms.

To investigate the benefits of the proposed logic family, we used our qPALACE toolsuite \cite{fourie2023results} to synthesize a 1024-point FFT with a 16-bit floating-point number system. The power values are derived from the average switching power of each element and the static power consumption in the distribution network. They are estimated using the qPA tool in the qPALACE toolsuite. In SCE, each JJ consumes about $I_c \Phi_0$ energy per switching event. The RSFQ library used is from our ColdFlux library, and the FPL library is available on GitHub \cite{Razmkhah2024}. We emphasize that these results should be interpreted as design-driven feasibility estimates, not experimental benchmarks, and are provided to motivate the potential benefits of developing an FPL-compatible fabrication process.

Table~\ref{tab:FFT_Comp} is the analytical projection derived from SPICE models and the gate-level synthesis from our toolsuite, demonstrating the comparison between CMOS, SFQ, and FPL technologies. We selected a 1024-point FFT with a 16-bit floating-point number system. The total memory, including feedback and twiddle factor, needed for such a structure is 65,536 bits. Here, the CMOS memory is SRAM, whereas for SFQ, we used DFF-based shift registers. FPL uses bistable vortex memory \cite{karamuftuoglu2024superconductor}, which is much more compact and power-efficient than DFFs. The transistor and JJ counts are just for the datapath and do not include memory. The power given includes only dynamic power, since the static power in SFQ and FPL depends on the power-delivery implementation and can approach zero with an AC power distribution network. The CMOS technology is a 12~nm FinFET, while for SFQ it is the MIT LL SFQ5ee process \cite{tolpygo2014fabrication}, and for FPL it is as described in this work. 
 
\begin{table}[htbp]
\centering
\caption{Projected performance comparison for a 1024-point floating-point FFT implemented using CMOS, SFQ, and FPL logic with the indicated fabrication technology. CMOS and SFQ values are based on reported experimental implementations, while FPL values are analytical projections derived from modeled device parameters and circuit-level SPICE simulations, as described in Section~\ref{Sec:SimMethod}. These results indicate trends in scalability rather than experimentally validated measurements. For SCE, a 300$\times$ cooling-power overhead is included. }
\label{tab:FFT_Comp}
\small
\begin{tabular}{|p{1cm}|p{1.4cm}|p{1.4cm}|p{1.2cm}|p{1cm}|}
\hline
\textbf{Tech. Set} & \textbf{Through-put (MOp/s)} & \textbf{Datapath FET/JJ} & \textbf{Memory type} & \textbf{Power (mW)}\\ \hline \hline
CMOS, 12~nm & 0.977 & 509,060 & RAM & 14.94 \\ \hline
SFQ, SFQ5ee & 15,625 & 1,296,120 & SR & 5.04  \\ \hline
SFQ, 0,stack & 15.625 & ~7,700,000 & SR &  3.36 \\ \hline
FPL, 0,$\pi$,stack & 39.062 & 4,536,420 & BVM & 2.4  \\ \hline
\end{tabular}
\vspace{-2em}
\end{table}

\begin{figure}[htpb]
\begin{center}
\includegraphics[width=0.9\linewidth]{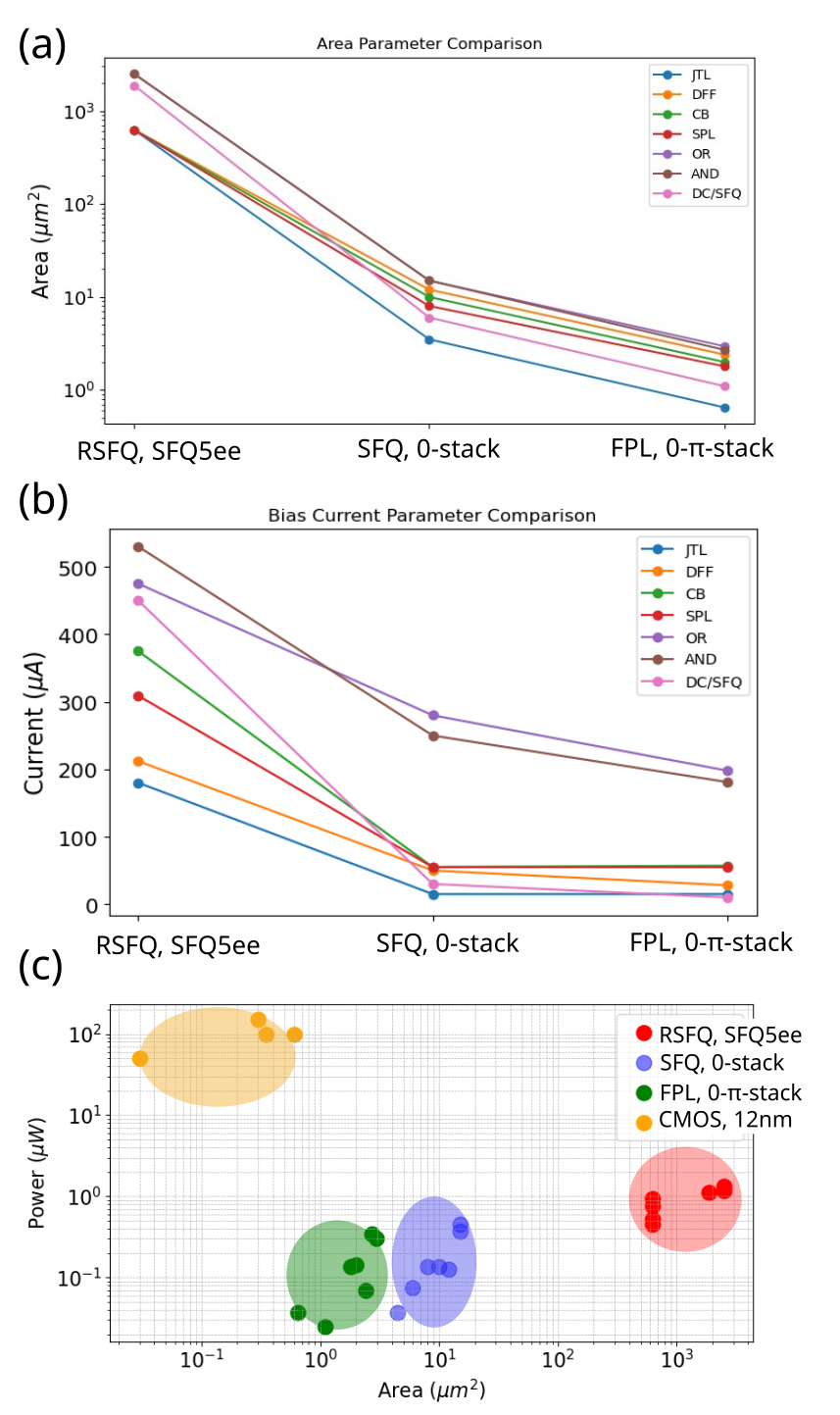}
\end{center}
\caption{Comparison of FFT components implemented using different technology sets: (a) Circuit area, (b) Supply current, and (c) power versus area, the power values are calculated with consideration of 300$\times$ cooling power overhead. }\label{fig:8}
\vspace{-1em}
\end{figure}

\section{Results}

textcolor{red}{It is important to emphasize that the FPL area and performance values reported in this section are analytical projections derived from modeled device parameters, layout-driven physical estimates based on target stacked-junction footprints, and circuit-level SPICE simulations. They represent feasibility estimates rather than measurements from experimentally fabricated chips. To bridge the gap between these projections and full device-level physical validation, ongoing experimental efforts are focused on developing robust, large-scale VLSI-capable fabrication processes for heterogeneous 0 and $\pi$ stacked junctions. }
Fig.~\ref{fig:8}(a) compares the area requirements for RSFQ, stacked-JJ, and FPL cells, showing that FPL achieves a reduction of at least 100$\times$ area compared to RSFQ. Furthermore, the supply current requirement is reduced fivefold between RSFQ and FPL, as plotted in Fig.~\ref{fig:8}(b). Implementing FPL with 0- and $\pi$-JJs improves integration and power efficiency, reduces circuit latency, increases throughput, and minimizes crosstalk and flux trapping. The higher circuit impedance also improves interfacing between FPL and CMOS circuits. Fig.~\ref{fig:8}(c) illustrates the various technologies in terms of power versus area. Although the CMOS power is high, it is still necessary for high-density memory.

In large-scale implementation, while the results in Table~\ref{tab:FFT_Comp} show that SFQ has a significant improvement over CMOS implementation in throughput and power, the current MIT LL SFQ5ee fabrication process can only support around 100~kJJ/\si{\centi\meter\squared} for logic and 200~kJJ/\si{\centi\meter\squared} for the implementation of shift register memory. Implementing the SDF FFT architecture with memory would require a chip area of 156~\si{\centi\meter\squared} (12.5~cm $\times$ 12.5~cm), which is impractical. For FPL with a density of 30~MJJ/\si{\centi\meter\squared}, the required chip area would be 1.4~\si{\centi\meter\squared} (1.2~cm $\times$ 1.2~cm), which is much more practical.

\section{Conclusion}
Scaling up the complexity of superconductor electronic logic circuits requires overcoming current limits in circuit density, fabrication, and architectural integration---factors that critically govern the commercial viability of this beyond-CMOS technology. In this work, we introduced the fast phase logic family to address key bottlenecks. By combining 0- and $\pi$-JJs with vertically stacked Josephson inductors (VSJJIs), FPL eliminates the need for bulky geometric wire inductors and external shunt resistors. When implemented with the proposed scalable NbTiN-based 0,$\pi$ stack fabrication process, FPL is projected to achieve a two-order-of-magnitude reduction in circuit area and a five-fold decrease in supply current compared to conventional RSFQ technology. Furthermore, the minimized inductive loops drastically reduce susceptibility to flux trapping and signal crosstalk.

Looking toward future improvements, the pathway to even higher integration levels will rely on 3D stacking of superconductor chips and heterogeneous integration with dense cryogenic memory, such as cryo-CMOS or bistable vortex memory (BVM). As fabrication improves, FPL has the potential to scale far beyond the 30~MJJ/cm\textsuperscript{2} density modeled in this work, enabling the monolithic integration of complex, deeply pipelined multi-core systems on practical chip areas. By closing the density gap that has historically hindered superconductor electronics, FPL provides a highly scalable, energy-efficient foundation to sustain the performance demands of next-generation, large-scale computing.

The architectural benefits, such as ultra-high-speed clocking, low-power continuous stream processing, and minimized memory overhead, position FPL as a candidate to replace traditional CMOS accelerators. As the thermal and power-delivery constraints in modern data centers become increasingly challenging, FPL-based superconductor accelerators offer a compelling alternative for large-scale computing required in real-time artificial intelligence inference, high-bandwidth signal processing, and general-purpose scientific computing.

\ack
\noindent
This work has been supported by DEVCOM under the FSDL: ColdPhase project (grant number W911NF2410317) and by the National Science Foundation (NSF) under the Expedition DISCoVER project (grant number 2124453).

The authors wish to acknowledge Mingye Li, Zeming Cheng, Jingkai Hong, and Ziyu Liu from the USC SPORT lab for their work on FFT synthesis results and superconductor EDA tool development.

\end{document}